\documentclass{astlb}
\usepackage{graphicx}
\usepackage{color}
\usepackage[hyperfootnotes=true]{hyperref}

\definecolor{darkblue}{rgb}{0,0,0.9}

\def\ps1{\emph{Pan-STARRS1}}

\begin{document}
\journalinfo{2022}{48}{12}{828}[838]
\title{Detection of AGNs and quasars having significant proper motions according to Gaia data within SRG/{\small e}Rosita X-Ray sources catalog}
\author{I.~M.~Khamitov\address{1,2,3}\email{irek\_khamitov@hotmail.com},
  I.~F.~Bikmaev\address{1,3},
  M.~R.~Gilfanov\address{4,5},
  R.~A.~Sunyaev\address{4,5},
  P.~S.~Medvedev\address{4},
  M.~A.~Gorbachev\address{1,3},
  E.~N.~Irtuganov\address{1,3}\\
$^1$\it{Kazan Federal University, Kazan, Russia\\}
$^2$\it{T\"UB\.ITAK National Observatory, Antalya, Turkey\\}
$^3$\it{Tatarstan Academy of Sciences, Kazan, Russia\\}
$^4$\it{Space Research Institute of Russian Academy of Sciences, Moscow, Russia\\}
$^5$\it{Max-Planck Institute for Astrophysics, Garching, Germany}
 }

\shortauthor{I. M. Khamitov et al.}

\shorttitle{Extragalactic sources in the SRG/eROSITA survey with proper motions}

\submitted{October 31, 2022}
 
\begin{abstract}

Based on a comparison of the SRG/eROSITA catalog of X-ray active stars and the Gaia catalog,  
a sample of 502 peculiar objects was obtained for which Gaia, on  one hand, detects statistically significant values of parallax or proper motion and, on the other hand, registers signs of the non zero source extent in the optical band. In the log ($F_X/F_{\rm opt}$)  -- (G-RP) color diagram  these objects are  separated from the balk  of X-ray active stars and are located in the region typical for the galaxies with active nuclei. According to the SIMBAD database, about $\sim 50\%$ of them are confirmed AGNs and galaxies with spectroscopically measured  redshifts, and only $\sim$1.4\% are confirmed Galactic objects. Spectroscopic observations of 19 unidentified objects on the RTT-150 telescope demonstrated, that 18 of them are AGNs at redshifts $\sim$0.01-0.3, and one object is a M star in our Galaxy. We discuss various scenarios explaining the nature of such peculiar objects.

%  \keywords{X-ray sources, active galaxy nuclei, optical observations, proper motions.}

  
\end{abstract}

\section{Introduction}

After more than two years of scanning the X-ray  sky eROSITA telescope  \citep{2021A&A...647A...1P} of the SRG Orbital X-ray  Observatory \citep{2021A&A...656A.132S} has detected an unprecedented number of X-ray sources and localized them with high positional accuracy. One of the directions of research with the obtained catalog of X-ray sources is to search for and study   stars in our Galaxy that are active in the X-ray. The GAIA  catalog \citep{GAIA, GAIAEDR3} is used for this purpose, which contains information on the parallaxes and 
proper motions of about an one and half billion  optical sources on the  sky. The combination of these  two catalogs makes it possible to form a catalog of X-ray active stars for further study. Examining the resulting catalog, we found a small number of objects with contradictory X-ray and optical characteristics that simultaneously pointed towards both extragalactic and Galactic nature of these sources.

Similar results were independently obtained earlier by \cite{LCAQ5} -- by comparing the large astrometric 
catalog of quasars (LQAC-5) and GAIA sources (eDR3) they identified  several dozen of extragalactic  sources for which GAIA measured statistically significant parallaxes and/or proper motions.
An example is the brightest (in the optical band) radio quasar 3C273, which has a measurable  proper motion \citep{2022A26A...667A.148G}. On the one hand, the measurement of proper motion for spectroscopically confirmed quasars may 
point to errors in one of the compared catalogs.   On the other hand, a possible explanation of such contradictory facts may be the displacement of the  photo-center of the galaxy or its active nucleus on the scale of GAIA observations, i.e. several years with significant positional measurements. This can occur, for example, due to the motion of jets in the  core or microlensing phenomena. Thus, based on new high-precision observations made for 4 extragalactic radio sources (3C 48, CTA 21, 1144+352, 1328+254) at VLBI in 2018-2021, a significant offsets of their positions from 20 to 130 milli arcsec at a time interval of over 2 decades \citep{2022MNRAS.512..874T} were found. The separation of such extragalactic sources from a sample of quasars and AGNs is extremely important for the task of building the fundamental GAIA coordinate system.

The present paper is dedicated to the study of such peculiar objects. 
In the section "Selection of sources"\ the methodology of selection of optical candidates of extragalactic sources  
with proper motions from the eROSITA catalog of  X-ray stars is described. In the section "Optical identification of the sample of extended sources"\ we perform  identification of the obtained sample by using  SIMBAD database. In the section "Optical spectroscopy of candidates"\,\ the results of spectral observations of 19 earlier
 non-identified sources carried out on the 1.5 m Russian-Turkish RTT-150 telescope during September -- October 2022 are presented. 
In the section "Analysis and discussion"\,\, possible scenarios that could explain the apparent proper motion of extragalactic sources are considered.

\section{Selection of sources}

\subsection{eROSITA telescope data}

In this work we used the SRG/eROSITA X-ray source catalog in the Eastern Galactic hemisphere,  the processing of data on which the Russian consortium of the eROSITA telescope is responsible for. The catalogue  was obtained by the X-ray catalog science working group of the Russian consortium of the  eROSITA telescope. A detailed description of the procedure for the detection and characterization of sources, astrometric correction, and validation of the catalog will be given in a dedicated paper. Here we present only the basic facts.
The catalog was built using the data obtained by the SRG/eROSITA telescope between December 2019 and February 2022. During this period eROSITA completed four  all-sky surveys  and performed a partially fifth survey covering $\approx 38\%$ of the sky.  Calibration of eROSITA telescope data, production of sky maps, detection of sources and their characterization 
were carried out using several components of the eSASS software developed by the German SRG/eROSITA consortium 
\citep{brunner2022} and the software developed at IKI RAS by the Russian SRG/eROSITA consortium. The data were processed using the results of ground-based calibrations and calibration observations, 
performed in October-November 2019. For further analysis the X-ray catalog  in the energy range of
0.3--2.3 keV will be used, filtered with the detection confidence threshold corresponding to $\approx 4 \sigma$ (likelihood threshold of 10).

\subsection{X-ray stars selection}

Galactic source candidates were selected based on a comparison with the GAIA satellite catalog. To  this end, the X-ray catalog was correlated with the GAIA eDR3 catalog and the eROSITA sources were selected so, that (i) in the 98\% error circle\footnote{ Typical values of the 98\% position error circle radius for eROSITA sources are $\approx$ 5--20 arcsec.} only one source from the GAIA catalog has been found, and (ii) 
for this source, the GAIA satellite measured the parallax or proper motion with a signal-to-noise ratio  S/N~$>~5$.  In determining the S/N of the proper motion measurement we  checked both the two components of proper motion in the equatorial coordinates, as well as the total proper motion of the source.  During the correlation of the eROSITA and GAIA catalogues, the positions of  optical sources were corrected for their proper motion in the cases when the signal-to-noise ratio of the proper motion measurement exceeded S/N~$>3$.

As a result of this selection, a catalog of likely  Galactic X-ray sources  was obtained, which included about $\sim$~$1.7\times 10^5$ sources. The vast majority of these objects are the stars  active in the X-ray range. The active binary stars, cataclysmic variables,  X-ray binaries and other less numerous classes of objects are contributing as well.

We do not consider issues of completeness and purity of the resulting catalogue, as they are not crucial 
for the purposes of this study. These issues will be discussed in subsequent publications. The main goal of this work is the study of a small group of  peculiar objects in the eROSITA catalog. At the same time we do not aim to make an exhaustive list of such objects, and their search procedure excludes chance coinsidences, at least from the point of view of their optical characteristics.

Nevertheless, let us make a few general remarks. First, we should note that the above analysis was performed 
using 98\% radius of localization of X-ray sources (localization errors of optical sources 
are many times smaller and are not important in this analysis). By the definition of 98\% error, 2\% of the objects must be 
outside this radius. It makes a relevant, but not decisive contribution to the completeness and purity of the resulting sample. 
A more important source of catalog impurity is random matches. As far as completeness with respect to 
Galactic objects, it is mainly determined by the fact that the X-ray 
sources with more than one GAIA object in the error circle were excluded.

For short, we will hereafter call the resulting sample of objects as "catalog of X-ray stars"\,, knowing that it also includes other types of Galactic sources and has limited completeness and purity.

\begin{figure}
  \centering

  \includegraphics[width=\columnwidth]{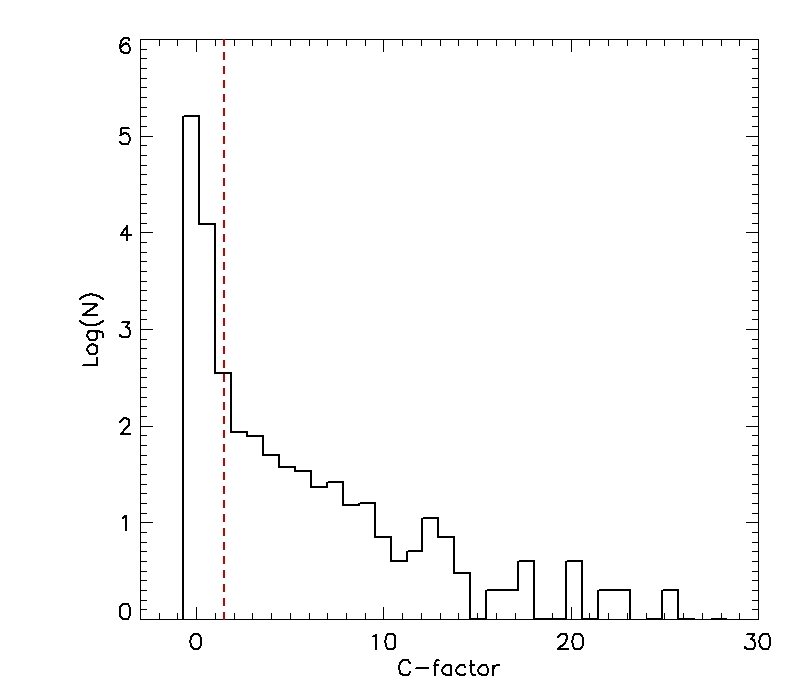}

  \hspace{1pt}
  \caption{The distribution of modified C*-factor values for the sources of the X-ray star catalog. 
For the selection of candidates for optically extended objects, values above 1.5 were used, i.e., objects to the right of the vertical dashed line in this graph.} 
  \label{fig:cfact}
  
\end{figure}

\begin{figure}
  \centering

  \includegraphics[width=\columnwidth]{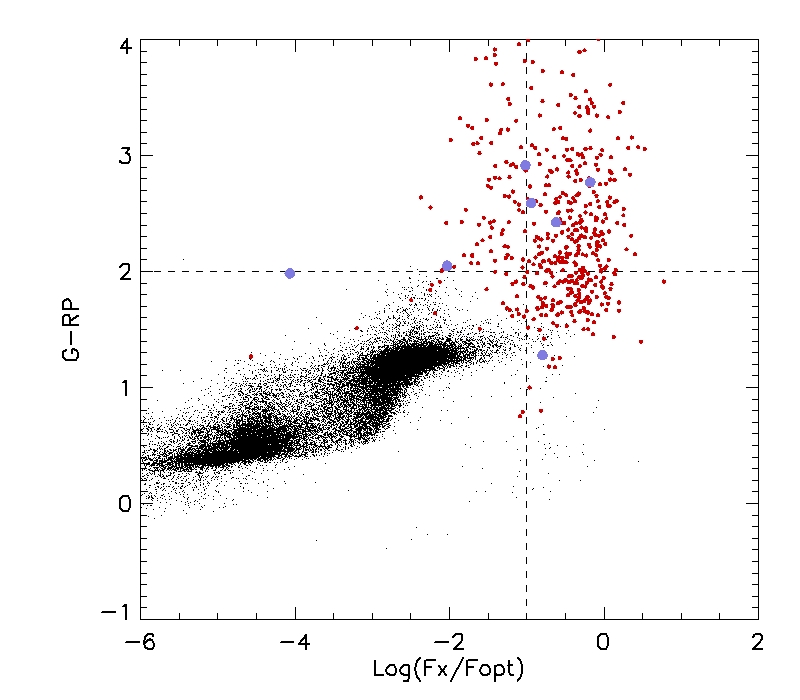}

  \hspace{1pt}
  \caption{Dependence of the G--RP color on the logarithm of the ratio of the X-ray flux measured from eROSITA data in the range 0.3--2.3~keV to the optical one, according to Gaia satellite measurements in the G-band, presented separately for a sample of close stars (black symbols) and a sample of extended objects (red symbols). Blue symbols indicate those of the candidates for optically extended objects that have been identified with Milky Way sources.   The dashed lines show the conditional boundaries separating stars from galaxies, AGNs, and quasars by color and with respect to $F_X/F_{\rm opt}$, see text. }
  \label{fig:FxFopt}
  
\end{figure}

\subsection{Search for extended optical sources in the catalog of X-ray stars}

In the GAIA eDR3 catalog, in addition to the integral flux in the broad photometric G-band, the fluxes 
in the BP and RP bands, which are (almost) two "halves"\ of the G band. Based on these, the parameter 
of color excess BP+RP (phot\_BP\_RP\_excess\_factor in GAIA catalog notation), calculated as the ratio 
of the sum of the fluxes in the BP and RP bands to the flux in the G band. In the literature, this value is sometimes denoted as C and 
is used to characterize the quality of GAIA satellite photometry and, in particular, as an indication of object extension  \citep{2021AA...649A...3R}. Despite the fact that the transmission in G-band almost completely covers the sum of 
BP and RP bands, due to the fact that the G-band uses narrower windows on the GAIA CCDs when transmitting data to the Earth, 
for extended or blended sources this C parameter will be greater than unity. For point sources, it is close to or slightly greater than unity, with some dependence on the shape of the object spectrum. In particular, because of the lower sensitivity of the G band in the red region of the spectrum as compared to the RP band, sources with emission detail can show significant value of the color excess parameter. Thus a corrected color excess factor C$^*$ was proposed, 
taking into account this effect \citep{2021AA...649A...3R}, which we will use further. The parameter C$^*$ is defined so that for point sources it is close to zero (as opposed to C). On Fig. \ref{fig:cfact}  
the distribution of C$^*$-factor values for the X-ray star catalog sources is shown.

Based on the catalog of X-ray stars, we have compiled two samples of objects. The first is all sources within 200 pc of the Sun, i.e., having a parallax  $> 5$ milli arcsec. Hereafter, we will call it 
the "close star sample". It characterizes the properties of X-ray stars with a sufficient degree of completness.  
The second sample is composed of sources with a corrected C$^*$-factor greater than 1.5, i.e., it is a sample of 
probably optically extended or blended objects (hereafter, the "extended object sample"). It includes 502 objects belonging to the catalog of X-ray stars. On Fig.~\ref{fig:FxFopt} these two samples are plotted on the plane of the logarithm of the ratio of X-ray flux to optical $F_X/F_{\rm opt}$ and the color G$-$RP. The ratio 
$F_X/F_{\rm opt}$ was calculated using the X-ray flux in the $0.3-2.3$ keV range and the optical flux in the G-band. The black dots show a sample of nearby stars, and the red dots show, respectively, the sources of the sample of extended objects. The horizontal dashed line shows the G$-$RP color boundary equal to 2, above which the  galaxies are mostly located \citep{2022arXiv220605681G}. The vertical dashed line $\log(F_X/F_{opt})=-1$ is the conditional boundary between the AGN and the stars \citep{belved23_pazh}. It is seen that in this diagram the optically extended sources are mainly located in the region typical to the active galaxies and quasars.

%\mg{Теоретически, большие значения цвета  G--RP и отношения $\log(F_X/F_{opt})$ у источников из выборки протяженных объектов на рис.\ref{fig:FxFopt} могли бы быть связаны с с более высоким межзвездным поглощением. Однако для искажение цвета G--RP на $\Delta_{\rm G-RP}\sim 1-2$ требуется значение избытка цвета $E(B-V)\sim ???$. Это нехарактерно высокие ??? значения для нашей Галактики. Согласно картам распределения пыли \citep{1998ApJ...500..525S}, такие значения $E(B-V)$ достигаются у менее чем $\la 5\%$ ??? обьектов из выборки протяженных источников. 
%} 

Thus, 502 X-ray sources were selected from the eROSITA catalog of X-ray stars, which are identified with GAIA peculiar objects. The peculiarity of the GAIA objects consists in the fact that, 
on the one side, they have statistically significant parallax or proper motion, which identifies them as objects in our Galaxy, and on the other side, according to GAIA's color excess parameter, 
they are probably optically extended objects. At the same time, these objects has an abnormally large measured G$-$RP color and/or high $F_X/F_{\rm opt}$ ratio, not characteristic for the stars (Fig.~\ref{fig:FxFopt}).
Subsequent sections of the paper are dedicated to the study of these unusual objects.

\begin{figure*}
  \centering

   \includegraphics[width=1.8\columnwidth]{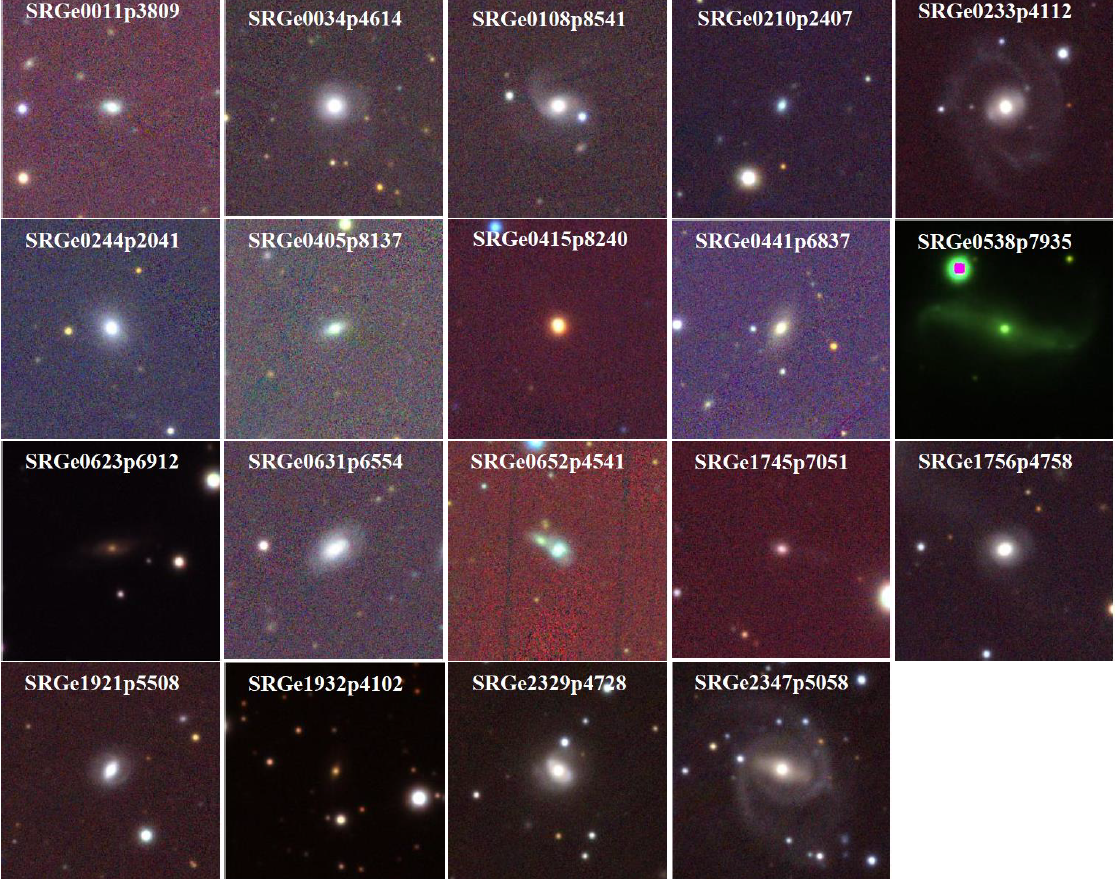}
  %\hspace{1pt}
  \caption{Visual inspection of the morphology of sources, carried out using deep images of the Pan-STARRS survey. The fields of 19 sources for which spectra were obtained on RTT-150 are presented. The size of each field is $60\times60$ arc sec. Only one source SRGeJ041510.2+824005 (figure identifier SRGeJ0415p8240) turned out to be an M-class star with a strong H$\alpha$ emission line. The other candidates were confirmed as extragalactic X-ray sources with redshifts from 0.016 to 0.272, active nuclei of galaxies of various types.}
  \label{fig:Inspection}
  
\end{figure*}

\section{OPTICAL identification of a sample of extended sources}

Identification of a sample of extended sources was performed using the SIMBAD database \footnote{https://simbad.cds.unistra.fr/simbad/sim-fcoo}. The search for matches was performed within a radius of 0.5 arcsec. We also found 11 matches with an accuracy of a few angular seconds, which we decided to keep in the list of matches for further study.

The results of the identification are listed in the Table \ref{tab:Simbad}. Totally 251 matches were found for the sources of a sample of extended objects as an extragalactic SIMBAD objects with spectroscopically measured redshifts. For 41 of them the proper motion modulus exceed 5 milli arcsec/year, and for four objects  -- 
more than 10 milli arcsec/year. 
Ten identified extragalactic objects with the largest proper motions are shown in the Table \ref{tab:AGNPM_result}.

It should be noted that according to the results of the identification with SIMBAD we found among the sample of extended sources 6 objects of Galactic nature. Among them, 4 sources are low-mass X-ray binaries with $C^*$-factor ranges from 4.5 to 8.5, one M-class star and one cataclysmic variable with $C^*$  is less than 2 (1.95 and 1.61, respectively). Moreover, the logarithm of $F_X/F_{\rm opt}$ exceeds --1 for CV, and for the M-star is about --4. The X-ray binaries are located near the M31 galaxy and the complicated background seems to be the reason for the large values of $C^*$.

For the sample of sources of extended objects absent in SIMBAD  a visual inspection of 
their morphology was performed using optical images of Pan-STARRS DR1 \citep{2016arXiv161205560C}; for this purpose web-interface of deep images  PanSTARRS-1 Image Access\footnote{https://ps1images.stsci.edu/cgi-bin/ps1cutouts} was used. 
The results of this process are shown in the bottom two rows of the Table \ref{tab:Simbad}.  The examples of the identification of extended sources are shown on Fig.~\ref{fig:Inspection}. Thus, a list of sources with an extended structure was compiled for further investigation of their nature by spectral observations.

\begin{table}[t]
%\vspace{6mm}
\caption{Results of identification analysis using the SIMBAD database  for the sources  in the sample of extended objects.} 
  \label{tab:Simbad}
%{{\footnotesize Table 1.} \footnotesize Результаты отождествления источников выборки протяженных объектов по Simbad \ik{Results of the identification of the sources of the sample of extended objects by Simbad}} \label{tab:Simbad} 
  \vskip 2mm
  \renewcommand{\arraystretch}{1.05}
  \renewcommand{\tabcolsep}{0.35cm}
 % \centering
  \footnotesize
 
      \begin{tabular}{lc}
    \hline

    Source type & 	$N$	 \\
            
\hline
Active galaxy nuclei   	    &     182 \\
The galaxies with measured z	    &     69 \\
Galaxies with unmeasured z	    &     39 \\
Galactic sources          &      6 \\
Unknown extended	        &     122 \\
Unknown blended      &     84 \\

 \hline
  
\end{tabular}
  
\end{table}

\section{Optical spectroscopy of candidates}

Optical spectroscopy was performed for 19  objects from the categories of "galaxies with unknown z"\ and 
"unknown extended sources"\ with G values brighter than $20^m$ and declinations above $\ge20$ degrees. Priority was given to objects with declinations of $\ge37$ degrees, since this program had the status of a pilot and observations were made during bright Moon times outside the main observational programs of optical support of the SRG observatory.

Spectral observations of these objects were carried out on the 1.5m Russian-Turkish RTT-150 telescope using the TFOSC\footnote{https://tug.tubitak.gov.tr/en/teleskoplar/rtt150-telescope-0} instrument. A CCD camera Andor iKon-L 936 \footnote{https://andor.oxinst.com/products/ikon-xl-and-ikon-large-ccd-series/ikon-l-936} was used with a $2048\times2048$ pixels BEX2-DD-9ZQ light detector  and a resolution element of 0.\arcsec326, cooled down to $-80^{\circ}$. 
The quantum efficiency of the CCD detector is  of 90\% or higher in the wavelength range from 4000\,\AA~to 8500\,\AA. Grism 15 was used as a disperser together with slit width of 0.134 mm (2.4 arcsec).
During September and October 2022 low-resolution spectra were obtained in the wavelength range from 3800\,\AA ~to 8800\,\AA ~ for 19 above mentioned candidates with a spectral resolution of 15\,\AA.
The spectral data were processed in a standard way using the \emph{IRAF}\footnote{http://iraf.noao.edu/} software, as well as with our own software using standard set of calibrations. The spectrophotometric calibration of the instrumental spectra was obtained 
using observations of spectrophotometric standards at the same zenith distance, as the investigated objects \citep{1990AJ.....99.1621O}.

\begin{figure}
  \centering
  \includegraphics[width=\columnwidth]{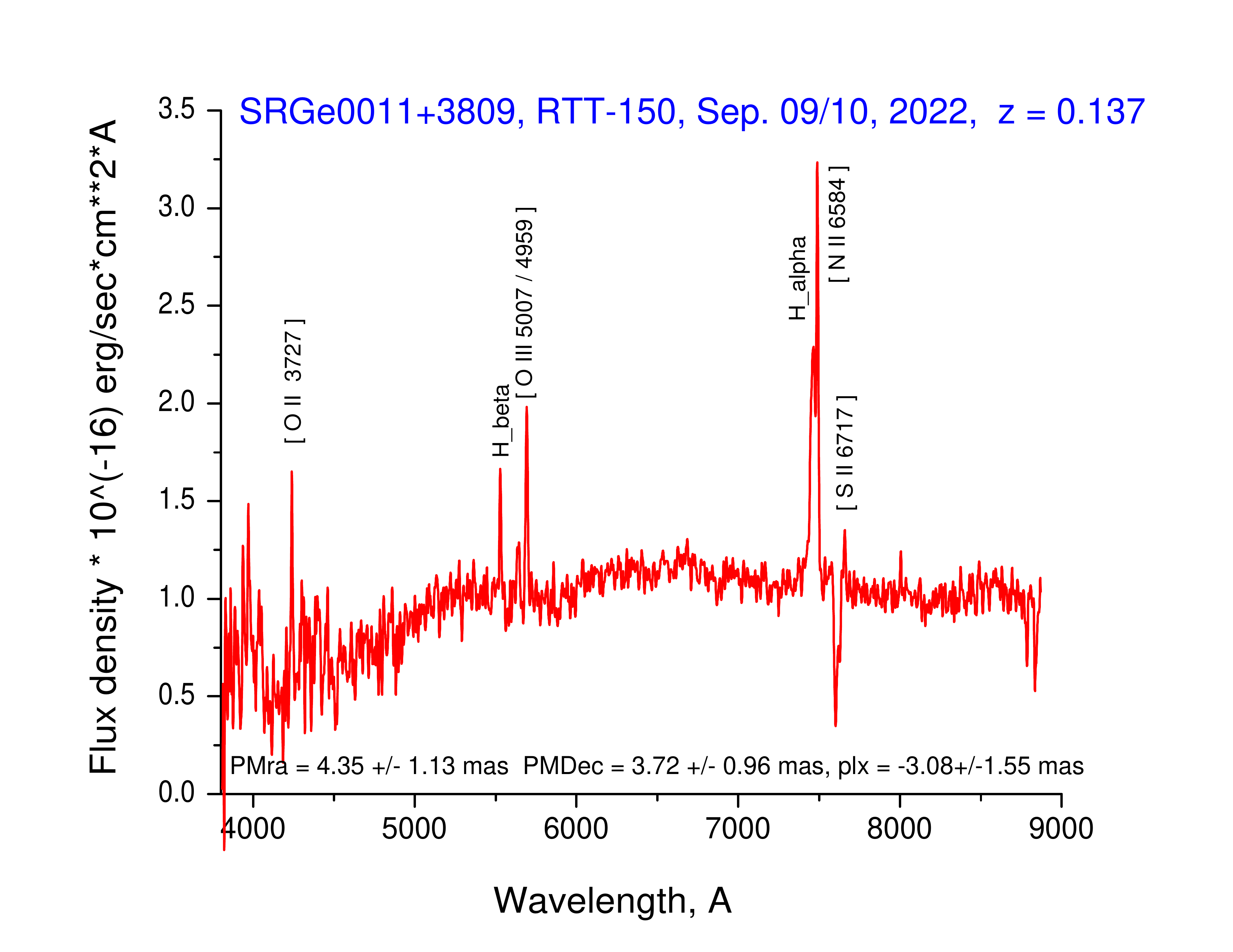}
  \hspace{1pt}
  \caption{The spectrum of SRGeJ001124.6+380935 of one of the 19 sources studied on RTT-150. The spectrum corresponds to the AGN type Sy2 at the redshift z = 0.137.}
 \label{fig:SRGe0011+3809sp}
  
\end{figure}

The results of spectral observations on RTT-150 are shown in Table \ref{tab:RTT150_result}.  
In Fig.~\ref{fig:SRGe0011+3809sp} the example of identification spectrum of SRGeJ001124.6+380935 object is shown. This candidate was  identified as an active galaxy nucleus of Sy2 type. Only one source of 19 -- SRGeJ041510.2+824005 -- was identified as an M-class star with a strong H$\alpha$ emission line. The remaining 18 candidates were confirmed as extragalactic sources. All of them turned out to be 
AGNs of various types with redshifts varying from 0.0155 to 0.272. The uncertainty of the redshift measurements is $\delta z/(1+z)\sim 0.001$.  The redshift measurement 
of the source SRGeJ024443.5+204136, with a rather general determination of its type as an extragalactic object, is also available in the LAMOST DR7 catalog \citep{2022ApJS..258....9W}, $z=0.050928\pm0.00004$, which agrees within the accuracy of the RTT-150 data. 

The program of spectral observations of the sample of extended objects which are accessible to observations on RTT-150 (DEC~$ >-30$\textdegree, $G<20^m$) will be continued. Objects fainter than {$G>20^m$} are planned for spectral observations with telescopes of larger apertures.

%\begin{landscape}
\begin{table*}
\caption{Identified extragalactic sources with spectroscopic redshift measurements that demonstrate the presence of significant apparent proper motions. The Table shows the 10 sources with the largest proper motions, in order of decreasing this value.}
\label{tab:AGNPM_result}
\vskip 2mm
\renewcommand{\arraystretch}{1.05}
\renewcommand{\tabcolsep}{0.2cm}
\linespread{1.3}
% \centering
\footnotesize
\tiny

\begin{tabular}{llcccccrll}
%\hrule
\hline 
eROSITA & GAIA eDR3 &  RA          & DEC         &   sep       & Gmag &      BP-RP &   $\mu$~~~~~     &        z       &       Type \\
(SRGe+) &           &              &             &   (arcsec)  &      &            &   (mas/yr)       &                &            \\
 
\hline

%SRGe1844p6248 & 18 44 26.31 & +62 48 29.8 & 18.97 & 1.58 & $  28.09 \pm 0.89 $ & 1.880 & [VV2006]J184426.2+624828(QSO)$^*$ \\
%\color{red}SRGe1735p2520 & 17 35 33.35 & +25 20 45.4 & 19.38 & 1.31 & $  17.33 \pm 0.28 $ & 0.01476 & 2MASXJ17353327+2520463(Galaxy) \\
%\color{red}SRGe1311p4635 & 13 11 18.54 & +46 35 02.3 & 19.45 & 1.27 & $  14.94 \pm 0.84 $ & 0.271342 & 2MASSJ13111853+4635024(QSO)$^*$ %\\
%SRGe2041m3811 & 20 41 13.45 & -38 11 37.5 & 19.05 & 0.99 & $  10.49 \pm 0.41 $ & 0.020204 & ESO341-4(Galaxy) \\
%SRGe1608p2957 & 16 08 51.07 & +29 57 15.0 & 19.91 & 1.62 & $   9.98 \pm 1.51 $ & 0.04849 & 2MASXJ16085109+2957144(Sy2) \\
%SRGe1105p5059 & 11 05 04.21 & +50 59 49.9 & 19.86 & 1.36 & $   9.43 \pm 1.30 $ & 0.11829 & LEDA2386530(Sy1)$^*$ \\
%SRGe1440p3327 & 14 40 25.84 & +33 27 02.6 & 20.82 & 1.46 & $   9.41 \pm 2.24 $ & 0.27474 & SDSSJ144025.84+332702.5(Sy1) \\
%SRGe0144p3140 & 01 44 17.27 & +31 40 03.3 & 20.09 & 1.81 & $   9.32 \pm 1.93 $ & 0.123625 & SDSSJ014417.27+314003.3(QSO) \\

%\color{red}SRGe1454p4645 & 14 54 25.48 & +46 45 24.1 & 18.99 & 1.29 & $   8.99 \pm 0.81 $ & 0.06914 & 2MASXJ14542548+4645239(Sy1) \\
%SRGe1356p2643 & 13 56 20.70 & +26 43 54.4 & 19.66 & 1.48 & $   8.86 \pm 1.54 $ & 0.06178 & LEDA1786612(Sy1) \\
 
J184426.6+624831 & 2157144511215355264 & 18 44 26.31 & +62 48 29.8 & 0.62 & 18.97 & 1.58 & $  28.09 \pm 0.89 $ & 1.880 & QSO$^{**}$ \\ 

J173533.5+252042$^{*}$ & 4593766796395321344 & 17 35 33.35 & +25 20 45.4 & 1.42 & 19.38 & 1.31 & $  17.33 \pm 0.28 $ & 0.01476 & galaxy \\

J131118.5+463502$^{*}$ & 1553988166345499136 & 13 11 18.54 & +46 35 02.3 & 0.27 & 19.45 & 1.27 & $  14.94 \pm 0.84 $ & 0.271342 & QSO$^{**}$ \\

J204113.5-381140 & 6682056384781697536 & 20 41 13.45 & -38 11 37.5 & 5.12 & 19.05 & 0.99 & $  10.49 \pm 0.41 $ & 0.020204 & galaxy \\

J160851.4+295719 & 1318789194505704192 & 16 08 51.07 & +29 57 15.0 & 0.18 & 19.91 & 1.62 & $   9.98 \pm 1.51 $ & 0.04849 & Sy2 \\

J110503.8+505951 & 839151745380126336 & 11 05 04.21 & +50 59 49.9 & 0.18 & 19.86 & 1.36 & $   9.43 \pm 1.30 $ & 0.11829 & Sy1$^{**}$ \\

J144026.2+332703 & 1286762448016026112 & 14 40 25.84 & +33 27 02.6 & 0.24 & 20.82 & 1.46 & $   9.41 \pm 2.24 $ & 0.27474 & Sy1 \\

J014417.5+314004 & 303683913296921728 & 01 44 17.27 & +31 40 03.3 & 0.13 & 20.09 & 1.81 & $   9.32 \pm 1.93 $ & 0.123625 & QSO \\

J145425.5+464525$^{*}$ & 1590336508929283200 & 14 54 25.48 & +46 45 24.1 & 0.23 & 18.99 & 1.29 & $   8.99 \pm 0.81 $ & 0.06914 & Sy1 \\

J135620.6+264356 & 1450823498570440832 & 13 56 20.70 & +26 43 54.4 & 0.19 & 19.66 & 1.48 & $   8.86 \pm 1.54 $ & 0.06178 & Sy1 \\
 
 \hline
\end{tabular}

$^{*}$ Sources with a ratio of the total displacement to the of the astrometric model discrepancy $(\mu \times 2.8)/\epsilon_i$ is greater than 5 (see "Analysis and discussion").

 $^{**}$ Sources of the large astrometric catalog of quasars LQAC-5.

\end{table*}

\section{ANALYSIS AND DISCUSSION}

\begin{table*}
\caption{ Results of spectral observations on RTT-150 of 19 sources from the sample of extended objects.}  
  \label{tab:RTT150_result}
  \vskip 2mm
  \renewcommand{\arraystretch}{1.05}
  \renewcommand{\tabcolsep}{0.2cm}
  \linespread{1.3}
 % \centering
  \footnotesize
\tiny

\begin{tabular}{llccccrrll}
%\hrule
\hline
eROSITA & GAIA eDR3 &	RA          & DEC         &   Gmag &      BP-RP &   $\mu_\alpha$~~~~~     & $\mu_\delta$~~~~~         &        z       &       Type \\
(SRGe+) &           &               &             &        &            &   (mas/yr)              & (mas/yr)                  &                &            \\ 

\hline

J001124.6+380935 & 2877771750682930176 & 00 11 24.46 & +38 09 33.7 &  19.92 &    1.14     & $ 4.35 \pm 1.13$ & $ -3.72 \pm 0.96$ &       0.137  &	AGN \\
J003447.2+461429 & 389161012394980352 & 00 34 47.72 & +46 14 29.3 &  18.87 &    1.38     & $ 2.37 \pm 0.39$ & $ -1.87 \pm 0.32$ &       0.1344  &	AGN    \\
J010812.2+854152 & 573881985223210368 & 01 08 11.86 & +85 41 50.7 &  19.39 &    1.38     & $ 0.33 \pm 0.60$ & $ -3.34 \pm 0.55$ &  	0.0772  &	Sy2    \\
J021049.7+240709 & 104209409979045376 & 02 10 49.38 & +24 07 06.7 &  19.34 &    1.35     & $-0.79 \pm 0.52$ & $ 2.593 \pm 0.47$ &   0.1437  &   Sy1    \\ 
J023309.6+411225$^{*}$ & 338399652915819648 & 02 33 09.64 & +41 12 22.5 &  19.16 &    1.34     & $ 4.68 \pm 0.85$ & $ -7.46 \pm 0.94$ &       0.062  &	Sy1 \\
J024443.5+204136 & 85364953204844416 & 02 44 43.31 & +20 41 38.6 &  18.68 &    1.33     & $-1.54 \pm 0.35$ & $  0.92 \pm 0.31$ &   0.0508  &  	Sy2 $^{**}$    \\
J040550.9+813716 & 569719612155006208 & 04 05 51.36 & +81 37 17.2 &  19.25 &    1.54     & $-0.87 \pm 0.31$ & $ -1.77 \pm 0.31$ &  	0.118   &      	Sy1    \\
J041510.2+824005 & 569939106459907328 & 04 15 10.62 & +82 40 11.8 &  16.76 &    3.17     & $20.88 \pm 0.15$ & $-36.98 \pm 0.11$ &               & 	M-star \\
J044110.7+683728 & 495873426235161344 & 04 41 11.00 & +68 37 29.3 &  19.78 &    1.72     & $ 3.32 \pm 0.38$ & $  1.99 \pm 0.65$ &    	0.1205  & 	AGN    \\
J053821.8+793515 & 553597546173225216 & 05 38 23.46 & +79 35 12.7 &  17.53 &    1.38     & $-2.21 \pm 0.40$ & $  3.66 \pm 0.46$ &    	0.0155	& 	AGN    \\
J062329.3+691238 & 1106940173151909888 & 06 23 29.08 & +69 12 32.6 &  19.93 &    1.68     & $-2.46 \pm 0.57$ & $ -3.82 \pm 0.73$ &    	0.0548  & 	AGN    \\
J063147.1+655440 & 1104230151864373888 & 06 31 47.43 & +65 54 42.5 &  19.36 &    1.29     & $-0.33 \pm 0.35$ & $ -1.94 \pm 0.39$ &  	0.1075	& 	AGN    \\
J065210.0+454141 & 954093346623632512 & 06 52 10.08 & +45 41 44.5 &  19.30 &    1.30     & $ 0.36 \pm 0.82$ & $ -3.75 \pm 0.72$ &    	0.1091  & 	Sy2    \\
J174514.8+705128 & 1639075140673511424 & 17 45 14.77 & +70 51 26.8 &  19.42 &    1.29     & $ 0.70 \pm 0.32$ & $ -2.17 \pm 0.41$ &    	0.272   & 	Sy1    \\
J175610.9+475824 & 1363075598726822912 & 17 56 10.81 & +47 58 24.8 &  18.67 &    1.38     & $-2.43 \pm 0.37$ & $ -0.53 \pm 0.35$ &    	0.0619	& 	Sy2 \\
J192156.1+550847 & 2140251786365419008 & 19 21 56.06 & +55 08 47.1 &  18.36 &    1.21     & $-3.01 \pm 0.23$ & $ -1.37 \pm 0.25$ &   	0.092   &      	Sy2    \\
J193203.8+410244 & 2053551343150878336 & 19 32 04.02 & +41 02 43.4 &  18.52 &    1.34     & $-0.93 \pm 0.18$ & $ -0.12 \pm 0.19$ &    	0.0835  &      	Sy1    \\
J232910.3+472800 & 1941757330160593536 & 23 29 10.44 & +47 28 01.3 &  18.55 &    1.47     & $-1.65 \pm 0.38$ & $  2.86 \pm 0.41$ &    	0.04	& 	Sy1    \\
J234726.2+505850 & 1944173403888702592 & 23 47 26.32 & +50 58 50.4 &  19.53 &    1.75     & $-2.77 \pm 0.62$ & $  4.24 \pm 0.59$ &       0.0621	& 	Sy1 \\

\hline
\end{tabular}

$^{*}$ Source with the ratio of the total displacement to the astrometric model dicrepancy parameter $(\mu \times 2.8)/\epsilon_i$ is greater than 5 (see "Analysis and discussion").

$^{**}$ The spectral measurement of the redshift of the source is also available in the LAMOST DR7 catalog 
\citep{2022ApJS..258....9W}, $z=0.050928\pm0.00004$.

\end{table*}

%\end{landscape}

%\label{sec:obs}

\subsection{Possible astrophysical explanations for the apparent proper motions of extragalactic objects}

The presence of apparent proper motions of extragalactic sources may be a result of object photocenter's position changes, which can take place for a number of reasons of astrophysical nature. The position of the photocenter of a galaxy with an active nucleus, measured by GAIA, is determined by the sum of the contributions of the radiation from the stellar component, accretion disk, nucleus, jets, and radiation from the medium interacting with the jets.  The phenomenon of the microlensing can also play a role.  The variability  of the different components which has a relative contributions to the registered brightness can leads to a time dependence of the optical coordinates of the AGN. The proper motion represents the average value of the photocenter position change over a period of 
more than 2.8 years, used in GAIA eDR3 data in the construction of astrometric solutions  \citep{GAIAEDR3}.  Consequently, the total displacement of the photocenters ($l$) of the studied objects is the value of the proper motion multiplied by the factor $\sim 3$.  Thus, to provide the observable apparent proper motion of AGN, a few components of radiation must be at the angular distances from several units 
to several tens of milli arcsec, and their brightness has to vary on a scale of $\sim 3$ years. Such angular distances correspond to linear sizes ranging from a few parsecs to several hundred parsecs, depending on the distance to the object.  These distances between the components can be achieved as a result of the motion of the jet and its interaction with the environment.  Indeed, in \citep{2019MNRAS.482.3023P}, using the previously detected offsets in the positions between the VLBI radio sources (mainly AGNs) and their optical components in GAIA  \citep{2017MNRAS.467L..71P}, it was obtained that the lines of significant offsets are parallel to the direction of the jets in 62\% of cases.  It is also noted that the parallelism of the proper motions vector to the direction of jets takes place much more often in the case of large values of the proper motions modulus.

For an object located at a redshift of $\sim$0.05--0.25, one millisecond of arc corresponds to 
linear size of $\sim$1--4 pc and, therefore, the velocities of the visible component displacements must be superluminal, $\sim$1--10$c$ in order to explain the observed proper motions. In the case of 
relativistic jets at velocities close to the speed of light, significant visible jet displacements are observed 
on the scale of a few years. For example, the Hubble Space Telescope measured the superluminal motion in relativistic jets in M87 nucleus on the order of 100 millisecond of arc per 5 years \citep{1999ApJ...520..621B}. In relativistic cases the displacement of the photocenter will be determined already by the character of motion of the jet itself.

The position of the photocenter can also be influenced by the cases when a star of our Galaxy is projected onto the AGN nucleus. 
In this case, the parallactic displacement of the photocenter should also be observed, and if the star is at the periphery of the 
of the Milky Way, the star's contribution will be only in the apparent proper motion. For such cases, in the spectra of AGN 
the spectrum of the star itself will also be present, which can be distinguished by spectral observations with a high signal-to-noise ratio. In addition, in order to isolate such cases one of the parameters of the GAIA eDR3 catalog -- the frequency of detection of multiple peaks when determining the image parameters
(ipd\_frac\_multi\_peak), can be used. This modulus is expressed as a percentage of detection of double peak in one-dimensional 
observation window to the total number of measurements \citep{GAIAEDR3}. The values of this parameter in our sample do not 
exceeds 10\%, except for the source SRGeJ184426.6+624831 with the highest proper motion value, which has a double image structure in more than half of the cases (53\%). Thus, 
the anomalous proper motion value of the quasar SRGeJ184426.6+624831 can be explained by the presence of a 
Milky Way star in the subsecond region of the source image. However, the absence of a significant parallax does not exclude 
the case of a double nucleus in the quasar as well.

\subsection{Noise characteristics in Gaia astrometric solutions}

\begin{figure*}
  \centering
  \includegraphics[width=\columnwidth]{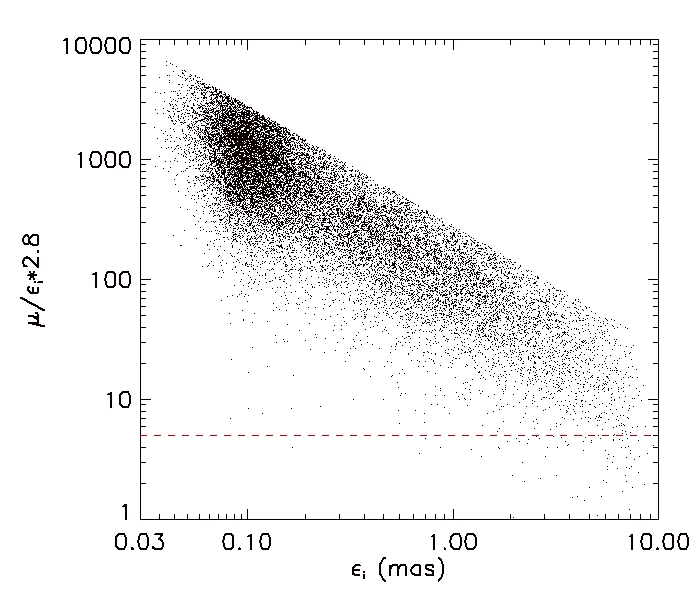}
  \includegraphics[width=\columnwidth]{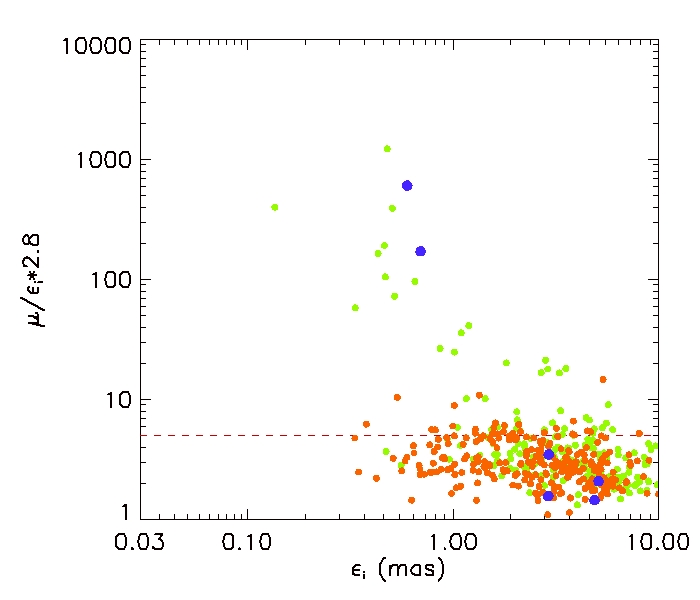}
  \hspace{1pt}
  \caption{The quality characteristics of the astrometric solutions for the catalog of nearby stars (left) and objects from the 
of the extended source sample (right). Both panels show only the objects for which the excess noise significance parameter $D>2$.  The scale is the same in both figures. The right panel shows objects whose extragalactic nature is confirmed by optical spectroscopy (orange circles), objects of our Galaxy (blue circles), and objects not identified in SIMBAD together with galaxies without spectral observations (light green circles).  The dotted line shows the level of the ratio $l/\epsilon_i=5.0$.}
  \label{fig:pm_epsi}
  
\end{figure*}

To evaluate the quality of the GAIA astrometric solutions and their agreement with the data, we used one of the parameters  
of the GAIA eDR3 catalog, the so-called source noise excess $\epsilon_i$ (astrometric\_excess\_noise). This value characterizes the discrepancy between the measured source positions and the 5-parameters astrometric model.  Positive values mean that the discrepancies are higher than the statistically expected values. The significance of the value $\epsilon_i$ itself is determined by the significance parameter D (astrometric\_excess\_noise\_sig). If $D>2$, the values of $\epsilon_i$ values are statistically significant. Excess noise characterizes various types of model and instrumental errors, which exceed errors of centroid determination. One would expect that sources with a complex and variable photocenter structure may have significant deviations from the standard 5-parameters 
astrometric model and correspondingly positive values of $\epsilon_i$. Fig.\ref{fig:pm_epsi} shows a plot of the ratio of the total photocenter displacement of the identified sources of the sample to the their source noise excess parameter ($l/\epsilon_i$) depending on the value of $\epsilon_i$ itself for close stars and extended sources.  The dashed line in the graph corresponds to $l/\epsilon_i$ equal to 5.

\begin{table*}
\caption{Known AGN and identified on the base of spectral data obtained on RTT-150 having the  ratio $l/\epsilon_i$ greater than 5. The sources are ordered in order of decreasing proper motion.}
\label{tab:PM_epsi_gt5}
\vskip 2mm
\renewcommand{\arraystretch}{1.05}
\renewcommand{\tabcolsep}{0.2cm}
\linespread{1.6}
% \centering
\footnotesize
\tiny

\begin{tabular}{lrccccrrcll}
\hline
 
eROSITA & GAIA eDR3 &   RA          & DEC         &   Gmag &   $\mu$       & $l/\epsilon_i$  & LOFAR     & VLASS      &       z       &       Type \\
(SRGe+) &          &               &             &        &   (mas/yr)    &         & (mJy)     & (mJy)        &               &            \\
 
\hline

%SRGe1735p2520 & 17 35 33.35 & +25 20 45.4 & 19.38 &  $  17.33 \pm 0.28 $ & $ 173.9 $  &          &          & 0.01476 & galaxy$^1$ \\
J131118.5+463502 & 1553988166345499136 & 13 11 18.54 & +46 35 02.3 & 19.45 &  $  14.94 \pm 0.84 $ & $   5.2 $  &  6.6     &          & 0.271342 & QSO$^1$ \\
J145425.5+464525 & 1590336508929283200 & 14 54 25.48 & +46 45 24.1 & 18.99 &  $   8.99 \pm 0.81 $ & $   5.4 $  &  3.7     & 1.71     & 0.06914 & Sy1$^2$ \\
J023309.6+411225 & 338399652915819648 & 02 33 09.64 & +41 12 22.5 & 19.16 &  $   8.81 \pm 1.24 $ & $   5.4 $  &          & 0.63     & 0.062 & Sy1$^3$ \\
J110241.8+420655 & 778252648175937920 & 11 02 41.47 & +42 06 51.9 & 19.35 &  $   7.02 \pm 0.90 $ & $   5.1 $  &          &          & 0.07498 & Sy1$^2$ \\
J212404.9-164149 & 6835161283006507520 & 21 24 04.81 & -16 41 48.1 & 19.17 &  $   7.01 \pm 1.09 $ & $   5.0 $  &          &          & 0.03588 & galaxy$^4$ \\
J004447.5+152910 & 2781106124341566976 & 00 44 47.34 & +15 29 11.9 & 20.22 &  $   6.51 \pm 1.17 $ & $   5.7 $  &          &          & 0.2272 & Sy1$^2$ \\
J233515.6-085729 & 2438794099819112960 & 23 35 16.07 & -08 57 23.5 & 20.04 &  $   5.23 \pm 1.04 $ & $   5.3 $  &          &          & 0.08566 & galaxy$^2$ \\
J192852.3-251642 & 6766602609845906176 & 19 28 52.31 & -25 16 38.7 & 19.17 &  $   5.20 \pm 0.55 $ & $  10.8 $ &          &          & 0.107730 & galaxy$^4$ \\
J144227.5+555848 & 1607566096654997760 & 14 42 27.61 & +55 58 46.4 & 18.61 &  $   4.61 \pm 0.33 $ & $   6.2 $ &  5.5     &   0.64   & 0.07689 & galaxy$^2$ \\
%SRGe1151p7108 & 11 51 01.67 & +71 08 31.8 & 20.19 &  $   4.52 \pm 1.10 $ & $  10.8 $ &          &          & 0.114 & Sy1$^6$ \\
J123740.2+611148 & 1580153004031345920 & 12 37 40.72 & +61 11 48.6 & 19.79 &  $   4.10 \pm 0.67 $ & $   5.4 $ &          &          & 0.18143 & Sy1$^2$ \\
J002936.8-173833 & 2367248023501811840 & 00 29 36.79 & -17 38 30.4 & 17.67 &  $   3.85 \pm 0.30 $ & $   6.0 $ &          &          & 0.05399 & galaxy$^4$ \\
J021257.5+140609 & 77274609208701824 & 02 12 57.60 & +14 06 10.2 & 18.24 &  $   3.66 \pm 0.41 $ & $   6.3 $ &          & 0.72     & 0.06172 & Sy1$^2$ \\
J143731.8+155549 & 1234415558406458752 & 14 37 31.70 & +15 55 47.6 & 18.21 &  $   3.64 \pm 0.45 $ & $   5.2 $ &          &          & 0.03701 & Sy1$^7$ \\
J113925.5+451345 & 773886418783213312 & 11 39 25.56 & +45 13 46.7 & 19.12 &  $   3.38 \pm 0.51 $ & $   5.4 $ &          & 0.67     & 0.1271 & Sy1$^2$ \\
J144924.5+321815 & 1283499819060669568 & 14 49 24.44 & +32 18 16.2 & 18.37 &  $   3.29 \pm 0.27 $ & $   5.8 $ &          & 4.61     & 0.058 & QSO$^6$ \\
J131447.1+260624 & 1447726002515875712 & 13 14 47.07 & +26 06 24.1 & 18.22 &  $   3.21 \pm 0.29 $ & $   8.8 $ &          &          & 0.07178 & Sy1$^2$ \\
J195456.5-062851 & 4196937111313151744 & 19 54 56.43 & -06 28 53.4 & 17.32 &  $   3.01 \pm 0.28 $ & $   5.1 $ &          &          & 0.029427 & galaxy$^4$ \\
J010816.3-113401 & 2469702471186911616 & 01 08 16.31 & -11 34 01.0 & 18.63 &  $   2.85 \pm 0.58 $ & $   5.1 $ &          &          & 0.04667 & galaxy$^4$ \\
%\color{red}SRGe0210p2407 & 02 10 49.38 & +24 07 06.7 & 19.34 &  $   2.71 \pm 0.60 $ & $   7.4 $ &          &          & 0.1437 & Sy1$^4$ \\
J070634.5+635057 & 1099887149653797120 & 07 06 34.82 & +63 50 56.1 & 18.00 &  $   2.58 \pm 0.27 $ & $   5.6 $ &          & 0.85     & 0.01425 & galaxy$^{7*}$ \\
J210221.6+105816 & 1756655887651319680 & 21 02 21.64 & +10 58 16.0 & 17.78 &  $   2.39 \pm 0.26 $ & $   5.2 $ &          & 1.31     & 0.02893 & Sy1$^7$ \\
J235601.9+073123 & 2746340185147938304 & 23 56 01.95 & +07 31 23.3 & 17.94 &  $   2.13 \pm 0.35 $ & $   6.0 $ &          &          & 0.040299 & Sy1$^8$ \\
%SRGe2156p1103 & 21 56 58.32 & +11 03 43.2 & 19.01 &  $   2.12 \pm 0.53 $ & $   6.7 $ &          &          & 0.1076 & Sy1$^7$ \\
%SRGe1437p2640 & 14 37 01.50 & +26 40 19.2 & 19.06 &  $   2.08 \pm 0.40 $ & $   8.1 $ &          & 3.66     & 0.2183 & QSO$^3$ \\
J014458.6-023200 & 2505449660085848576 & 01 44 58.56 & -02 31 59.0 & 17.73 &  $   1.98 \pm 0.24 $ & $  10.3 $ &          &          & 0.09573 & galaxy$^4$ \\
%SRGe0237p1938 & 02 37 59.99 & +19 38 11.8 & 18.75 &  $   1.68 \pm 0.44 $ & $   8.0 $ &          &          & 0.034108 & galaxy$^9$ \\
J164313.8+095416 & 4446025560705238656 & 16 43 13.78 & +09 54 16.2 & 17.33 &  $   1.63 \pm 0.17 $ & $   5.5 $ &          &  0.76    & 0.04727 & galaxy$^7$ \\
J090436.8+553603 & 1035985561071454080 & 09 04 36.95 & +55 36 02.7 & 17.57 &  $   1.57 \pm 0.15 $ & $   5.6 $ &          &          & 0.03724 & Sy1$^2$ \\
J224311.1+032804 & 2704625950939055488 & 22 43 11.02 & +03 28 04.8 & 17.15 &  $   0.84 \pm 0.14 $ & $   6.2 $ &          &  2.41    & 0.03913 & galaxy$^{9}$ \\

\hline
\end{tabular}
 $^*$ For this source, in addition to the optical component, two radio components were detected at distances of 0.5 and 0.54 arcsec from the optical component.

 \textbf{Note.} The LOFAR and VLASS measurements were made in the frequency ranges 120 -- 168~MHz and 2 -- 4~GHz, respectively, and the values refer to the $S_{\rm peak}$ peak flux with a significance greater than $5\sigma$.
 Redshift measurements and source identifications were used from the following catalogs:  
1 -- SDSS DR7 \citep{2009ApJS..182..543A}, 2 -- SDSS DR9 \citep{2012ApJS..203...21A}, 
3 -- RTT-150 (this work), 4 -- 6dFGS \citep{2009MNRAS.399..683J}, 
5 -- SDSS DR11/12 \citep{2015ApJS..219...12A}, 6 -- LAMOST \citep{2018AJ....155..189D}, 
7 -- UZC 
\citep{1999PASP..111..438F}, 8 -- ALFALFA \citep{2018ApJ...861...49H} and 9 -- 2MRS \citep{2012ApJS..199...26H}.

\end{table*}

In Fig.~\ref{fig:pm_epsi} it is seen that the stars have a high value of signal-to-noise ratio: for them the measured 
proper motion $\mu$ exceeds the excess astrometric noise $\epsilon_i$ by $\sim$1--2 orders of magnitude.  At the same time, for extended objects, the opposite picture is observed: the total displacement in the sky $l$ for for most objects is only 2 to 3 times greater than the astrometric noise $\epsilon_i$. This may indicate some irregularity in the movement of the photometric center of the extended objects compared to the prediction of the astrometric model. The small number of objects in the catalog of extended sources with high values of signal-to-noise ratio, are probably real galactic sources with proper motions. Note that the M-class star identified by the RTT-150 observations has a ratio $(\mu \times 2.8)/\epsilon_i\sim 200$.

In Table \ref{tab:PM_epsi_gt5} the 25 sources lying above this line are shown. Remarkably, that only 3 of the sources in Table  \ref{tab:AGNPM_result} with the largest proper motions have 
$l/\epsilon_i$ values above five.  Among the sources listed in Table \ref{tab:PM_epsi_gt5} as galaxy, according to the classification from SIMBAD, only SRGeJ192852.3-251642 has an optical spectrum, from the 6dFGS survey \citep{2009MNRAS.399..683J} obtained in 2003, corresponding to spectrum of elliptical galaxy with [K, H] Ca, G, Mg, Na absorption lines at the redshift $z=$0.10773 and with the complete absence of emission lines typical for AGN. There are significant radio components for 11 sources, either in the public part of the LoTSS survey  
in the 120--168 MHz range \citep{2019AA...622A...1S}, or in the VLASS northern sky survey in the 
2--4 GHz \citep{2021ApJS..255...30G}, which is 44\% of the total number of 25 sources. These are mostly Seyfert 
Type 1 galaxies and quasars. Among the 4 sources that are in the HETDEX spring field of the LoTSS survey, with a size of 
424 square degrees, only SRGeJ113925.5+451345 has no significant radio signal detected. 
Robust measurements of the radio component flux density ($S_{\rm peak} >3$ mJy/beam) from VLASS data are detected for the quasar SRGeJ144924.5+321815. It should be noted that the brightness variations of AGNs have a stochastic behavior, and the registered proper motions due to changes in the photocenter associated with physical processes in these systems may change direction in the future, or completely disappear.

In conclusion, we note that in this work we used the ready-to-use GAIA astrometric solutions published 
in the EDR3 catalog.  In choosing one or another scenario to explain the significant magnitudes of the apparent proper motion of AGN, it is necessary to analyze the traces of objects in the sky using the individual positions of sources measured by GAIA at different epochs. Such an analysis is planned to be presented in future publications.  Spectrophotometric observations of the objects, which will be continued on RTT-150 and other telescopes, also play a critical role.

\section{Conclusion}

Using Gaia measurements of proper motions and the value of the color excess factor, 
we have identified a small group of peculiar objects in  the SRG/eROSITA catalog of X-ray stars. 
Their peculiarity according to Gaia data is determined by the fact that, on  one hand, they have statistically significant 
measurements of parallax and/or proper motion, and on the other hand, the large value of color excess $C^*$ 
are indicative of their non-zero angular extent. The sample includes 502 such objects. On the  log ($F_X/F_{\rm opt}$) -- color (G--RP) diagram, they are located outside  the bulk population of stars, in a region more typical to a galaxies and AGNs. According to SIMBAD database, 251 sources are extragalactic and have redshifts measured by spectroscopic methods, 6 sources are known galactic objects, and 206 objects are not present in SIMBAD. The optical extension for 122 of the latter is confirmed visually using the Pan-STARRS data.

For 19 objects not presented in the SIMBAD database we 
performed low-resolution spectral observations on the 1.5-m telescope RTT-150 in September - October 2022. We found 18 previously unknown AGNs with redshifts ranging from 0.0155 to 0.272. On the other hand they all have proper motions according to GAIA measurements in the range from 0.9 up to 8.8 milli arcsec/yr. One source turned out to be an M-star in the Milky Way  with a strong H$\alpha$ emission line. 
Together with the six Galactic sources identified in SIMBAD, the fraction of confirmed Galactic sources in our sample is $\approx 1.4 \%$.   

The catalog of extended sources with proper motions will be published in a subsequent paper. The catalog can be made available to interested scientists studying such objects prior to publication upon a request to the first author of the paper.

\acknowledgements

This work is based on observations with the eROSITA
telescope onboard the SRG observatory. The SRG observatory was built by Roskosmos in the interests of the Russian Academy of Sciences represented by its Space Research Institute (IKI) within the framework of the Russian Federal Space Program, with the participation of the Deutsches Zentrum fuer Luft- und Raumfahrt (DLR). The
SRG/eROSITA X-ray telescope was built by a consortium of German Institutes led by MPE, and supported by DLR. The SRG spacecraft was designed, built, launched,
and is operated by the Lavochkin Association and its subcontractors. The science data are downlinked via the
Deep Space Network Antennae in Bear Lakes, Ussurijsk, and Baykonur, funded by Roskosmos. The eROSITA data used in this work were processed using the eSASS
software system developed by the German eROSITA consortium and the proprietary data reduction and analysis software developed by the Russian eROSITA Consortium. This research has made use of the SIMBAD database, operated at CDS, Strasbourg, France.
The Pan-STARRS1 Surveys (PS1) and the PS1 public science archive have been made possible through contributions by the Institute for Astronomy, the University of Hawaii, the Pan-STARRS Project Office, the Max-Planck Society and its participating institutes, the Max Planck Institute for Astronomy, Heidelberg and the Max Planck Institute for Extraterrestrial Physics, Garching, The Johns Hopkins University, Durham University, the University of Edinburgh, the Queen's University Belfast, the Harvard-Smithsonian Center for Astrophysics, the Las Cumbres Observatory Global Telescope Network Incorporated, the National Central University of Taiwan, the Space Telescope Science Institute, the National Aeronautics and Space Administration under Grant No. NNX08AR22G issued through the Planetary Science Division of the NASA Science Mission Directorate, the National Science Foundation Grant No. AST-1238877, the University of Maryland, Eotvos Lorand University (ELTE), the Los Alamos National Laboratory, and the Gordon and Betty Moore Foundation.
The work of I.M. Khamitov, I.F. Bikmaev, M.A. Gorbachev, and E.N. Irtuganov was supported by the subsidy N 671-2020-0052 of the RF Ministry of Education and Science, allocated to Kazan Federal University to fulfill the state task in the field of scientific activities. M.R. Gilfanov, P.S. Medvedev, and R.A. Sunyaev thank the support of the RNF grant 21-12-00343.
The authors are grateful to 
T\"UB\.{I}TAK, IKI, KFU, and the Academy of Sciences of the Tatarstan Republic  for partial support in use of RTT-150 (Russian-Turkish 1.5-m telescope in Antalya).

\bibliographystyle{astl}
\bibliography{ero_agn_pm}

\end{document}